# Tera-Hertz Counterparts to Fast Radio Bursts from Coherent Cherenkov Radiation by Tilted Bunches

Ze-Nan Liu[1,2,3], Wei-Yang Wang[4], Yu-Chen Huang[1,2], and Zi-Gao Dai[1,2]
[1] Department of Astronomy, University of Science and Technology of China, Hefei 230026, People's Republic of China; daizg@ustc.edu.cn
[2] School of Astronomy and Space Science, University of Science and Technology of China, Hefei 230026, People's Republic of China
[3] School of Astronomy and Space Science, Nanjing University, Nanjing 210023, People's Republic of China
[4] School of Astronomy and Space Science, University of Chinese Academy of Sciences, Beijing 100049, People's Republic of China
*Received 2025 April 22; revised 2025 June 26; accepted 2025 June 30; published 2025 August 8*

## Abstract

Fast radio bursts (FRBs) are millisecond-duration extragalactic transients characterized by ultrahigh brightness temperatures, suggesting coherent emission mechanisms in extreme astrophysical processes. In this paper, we extend the bunched coherent Cherenkov radiation (CChR) framework by incorporating bunch inclination and geometric configuration parameters, enabling it to more rigorously model FRB emission and tera-Hertz (THz) emission from magnetars. When relativistic bunches are injected into the magnetized plasma of a magnetar's magnetosphere at the Cherenkov angle, their emitted waves achieve phase coherence through constructive interference. Furthermore, the three-dimensional geometry of the bunches plays a crucial role in influencing the coherence of the radiation. Within the framework of CChR, we predict the existence of THz emission counterparts associated with FRBs and explain the observed characteristics of the THz-emitting magnetar SGR J1745-2900. Detections of such counterparts by upgraded millimeter telescopes (e.g., Atacama Large Millimeter/submillimeter Array, IRAM) would be expected to provide new insights into the potential physical connection between FRBs and magnetars.

*Unified Astronomy Thesaurus concepts:* Radio bursts (1339); Radio transient sources (2008); Magnetars (992); Radiative processes (2055)

## 1. Introduction

Fast radio bursts (FRBs) are a class of short-duration radio transients, lasting just a few milliseconds, characterized by extremely high brightness temperatures and significant dispersion measures (D. R. Lorimer et al. 2007). Despite extensive observational and theoretical efforts, the physical mechanisms responsible for their generation and emission remain elusive (B. Zhang 2020; D. Xiao et al. 2021). Recent studies by C. D. Bochenek et al. (2020) and CHIME/FRB Collaboration et al. (2020) have reported the detection of a luminous FRB-like burst originating from the Galactic magnetar SGR J1935+2154. This discovery provides compelling evidence that magnetars may serve as a viable progenitor for at least some FRBs. The high brightness temperature of FRBs strongly suggests that their emission must originate from a coherent radiation mechanism (D. R. Lorimer et al. 2007; B. Zhang 2020), where charged particles act collectively and organized rather than radiating independently. Several leading radiation mechanisms have been proposed to explain this phenomenon, including coherent emission by bunches, plasma masers, and vacuum masers, etc. (D. B. Melrose 2017). The radiation mechanisms of bunched coherent emission within the magnetosphere of a magnetar can be primarily categorized into the following three types: coherent curvature radiation (P. Kumar et al. 2017; W. Lu & P. Kumar 2018; Y.-P. Yang & B. Zhang 2018; W.-Y. Wang et al. 2022; Z.-N. Liu et al. 2023b, 2024), coherent inverse Compton scattering (B. Zhang 2022; Y. Qu & B. Zhang 2024), and coherent Cherenkov radiation (CChR; Z.-N. Liu et al. 2023a).

Cherenkov radiation occurs when relativistic charged particles move through a medium at a velocity exceeding the medium's phase velocity of light (J. V. Jelley 1955; P. A. Cherenkov 1958; J. D. Jackson 1998). As the field of "superluminal" particles propagates through the medium, it accelerates electrons that generate the secondary wave, making CChR a collective effect of particle interactions. The CChR mechanism has been investigated to explain various astrophysical (G. A. Askaryan 1962; H. Kolbenstvedt 1977; J. Alvarez-Muñiz et al. 2012; Z.-N. Liu et al. 2023a) and experimental physics (S. Y. Gogolev & A. P. Potylitsyn 2019; Y. Tadenuma et al. 2022) phenomena. Notably, Z.-N. Liu et al. (2023a) have introduced a novel radiation mechanism for FRBs, proposing that some relativistic particles (bunches) are emitted from the polar cap of a magnetar, traveling along magnetic field lines through a relativistic electron–positron plasma generated by the pair cascade process, where its interaction with the plasma gives rise to CChR along the particle trajectories. Interestingly, experimentalists have proposed a novel method for generating coherent tera-Hertz (THz) Cherenkov radiation by precisely controlling the tilt of the bunches (i.e., the geometric configuration of the bunch is tilted relative to the direction of motion; Y. Tadenuma et al. 2022). They achieved this by utilizing a 4.8 MeV electron bunch from a photocathode radio-frequency gun and adjusting the angle through a radio-frequency transverse deflecting cavity. Furthermore, S. Y. Gogolev & A. P. Potylitsyn (2019) simulated the characteristics of CChR and predicted a pronounced azimuthal asymmetry in the CChR emitted from the tilted bunch. Therefore, the inclination of the bunches and their three-dimensional geometric configuration will significantly influence the coherence of CChR.

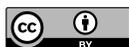

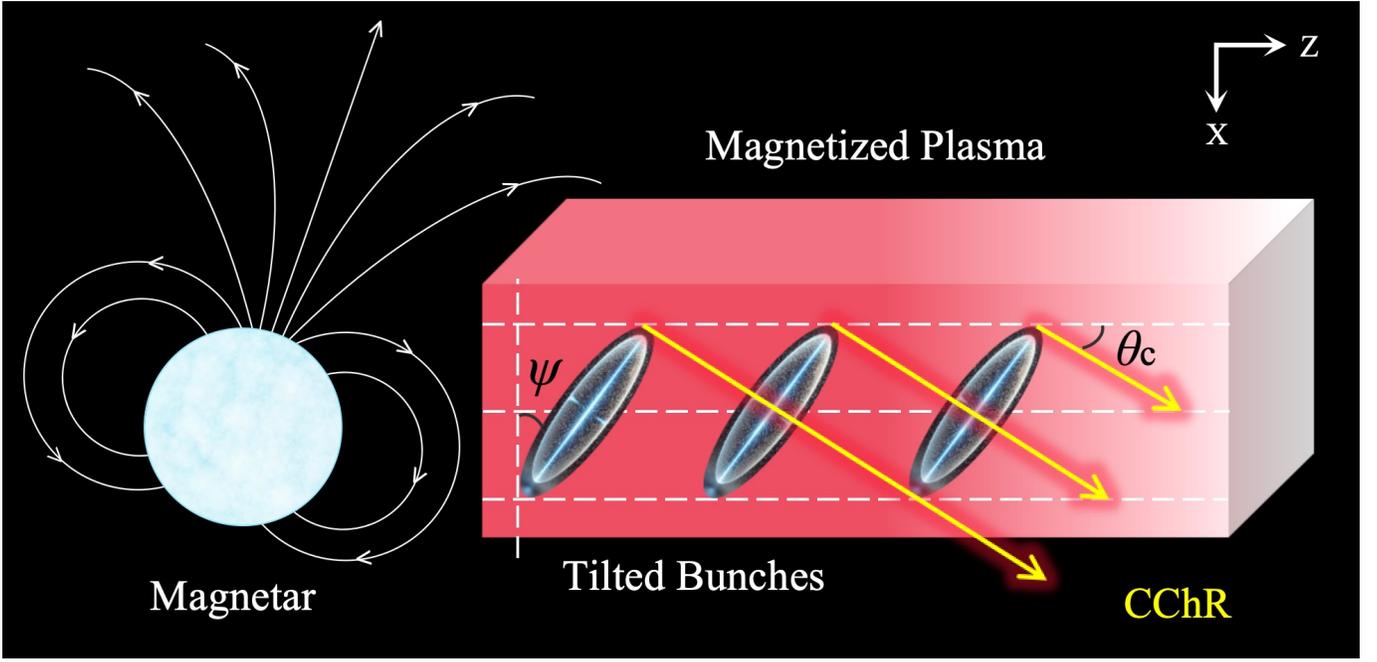

**Figure 1.** CChR from tilted bunches within the magnetosphere of a magnetar. The z-axis direction represents the velocity direction of the relativistic particles. The yellow arrows denote the emission direction of CChR, $\theta_c$ represents the Cherenkov radiation angle, and $\psi$ indicates the inclination angle of the bunch.

In this paper, we incorporate the inclination and geometric configuration of the bunches into the framework of CChR to investigate in greater depth its potential applications to FRBs and THz-emitting magnetars. The paper is organized as follows. In Section 2, we first study the CChR from a tilted bunch within the magnetosphere of a magnetar. Coherent THz Cherenkov radiation from FRBs and THz-emitting magnetars is discussed in Section 3. The results are discussed and summarized in Section 4. The convention $Q_x = Q/10^x$ in cgs units is used throughout this paper.

## 2. CChR from A Tilted Bunch

The coherence of CChR is strongly influenced by the inclination and geometric configuration of the bunches (S. Y. Gogolev & A. P. Potylitsyn 2019; Y. Tadenuma et al. 2022). As shown in Figure 1, we present the CChR from the interaction between tilted bunches and magnetized plasma within the magnetosphere of a magnetar. The Cherenkov angle $\theta_c$, defined as the angle between the radiation direction and the particle's motion, is given by $\cos\theta_c = c/n_r v$, where $n_r$ is the refractive index and $c/n_r$ is the speed of light in the medium. From the above equation, it is evident that the direction of Cherenkov radiation deviates from the velocity direction of the relativistic particles. When the particle beam is tilted, the charge distribution takes on an asymmetric Gaussian shape, with its specific form depending on the tilt angle $\psi$. Considering the tilt angle $\psi$ of the bunch as an additional parameter introduces another degree of freedom for the bunch (S. Y. Gogolev & A. P. Potylitsyn 2019; Y. Tadenuma et al. 2022). When the bunch is injected into the magnetic plasma medium at a precisely constrained tilt angle that matches the Cherenkov angle, the emission emitted from any location within the bunch can overlap coherently with the same phase.

Additionally, the three-dimensional geometry of the bunches can also affect the coherence of the radiation.

To calculate the spectral-angular distribution of radiation emitted by bunches, it is essential to first determine the refractive index of the medium. We consider a plasma composed of fully separated charges immersed in a magnetic field $B$. Plasma particles travel parallel to the direction of the magnetic field with a velocity $\beta_p$ and energy $\gamma_p$. In the comoving frame of the plasma, the dispersion equation for an electromagnetic wave of frequency $\omega'$, propagating at an angle $\theta'$ relative to the magnetic field direction, is given by the following expression (H. Heintzmann et al. 1975; Z.-N. Liu et al. 2023a):

$$\frac{\omega'^2 \omega_B^2}{\omega_p^4}(n_r'^2 - 1)\left[n_r'^2 \cos^2\theta' - 1 - (n_r'^2 - 1)\frac{\omega'^2}{\omega_p^2}\gamma_p^2 \Gamma\right]$$
$$= \Gamma\left(1 - \frac{\omega'^2}{\omega_p^2}\gamma_p^2 \Gamma\right)\left[1 + (n_r'^2 - 1)\frac{\omega'^2}{\omega_p^2}\right]^2,$$
(1)

where $m_e$ is the mass of the electron, $\Gamma = (1 - n_r'\beta_p\cos\theta')^2$, the relativistic electron gyration frequency $\omega_B = eB/\gamma_p m_e c$, and the relativistic plasma frequency $\omega_p = (4\pi n_e' e^2/\gamma_p m_e)^{1/2}$. For a magnetar characterized by a rotation period $P$ and a surface magnetic field $B_{s,15}$, the Goldreich–Julian charge number density (P. Goldreich & W. H. Julian 1969; M. A. Ruderman & P. G. Sutherland 1975) is given by $n_{GJ} \simeq 6.9 \times 10^7 \text{cm}^{-3} B_{s,15} P_0^{-1} R_8^{-3}$. For the detailed calculation of the refractive index in a magnetized plasma, one can refer to Z.-N. Liu et al. (2023a). In the Appendix, we provide a comprehensive derivation of the spectral-angular distribution of radiation emitted by bunches traveling through the magnetic plasma. For the case of CChR generated by a tilted bunch traveling through the magnetized plasma within the magnetosphere of a magnetar, our computational results are presented in Figure 2 and analyzed in detail below.

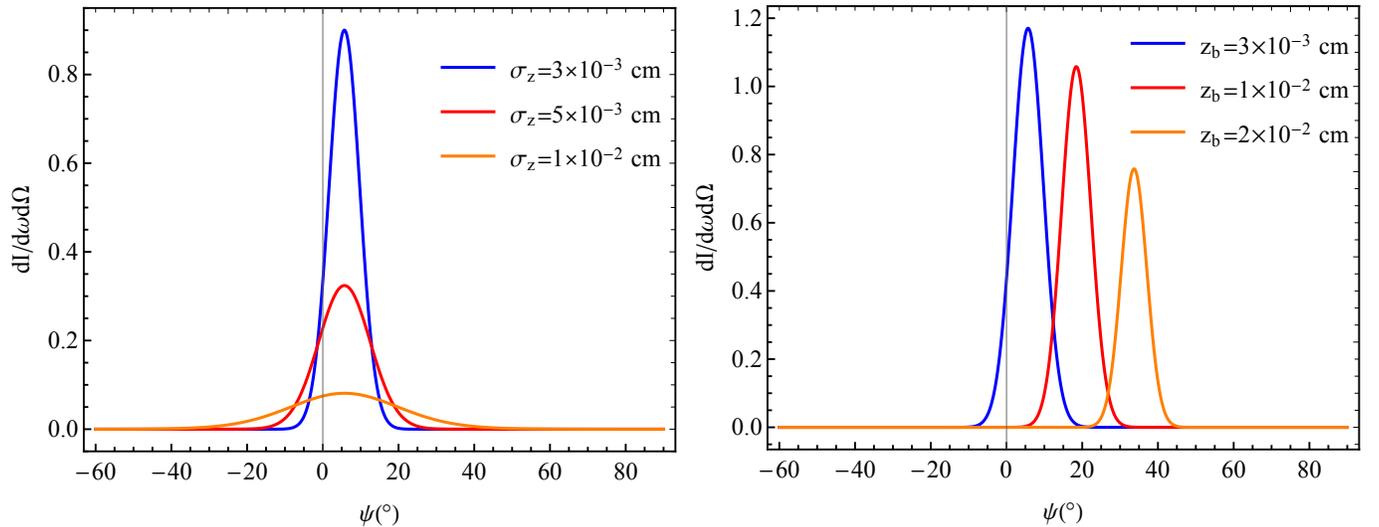

**Figure 2.** The spectral-angular distribution of radiation emitted by bunches. The left and right panels correspond to different values of $\sigma_z$ (the bunch size in the $z$-direction) and different $z_b$ (the position of the bunches along the $z$-axis), respectively. The following physical parameters are employed in our calculation: $\gamma_e = 10^{2.5}$, $\gamma_p = 10^2$, $P = 1$ s, $B_s = 10^{15}$ G, $\xi = 100$, $R = 10^8$ cm, and $x_b = y_b = \sigma_x = \sigma_y = 3 \times 10^{-2}$ cm.

The CChR emitted by bunches is fundamentally determined by the interplay of spatial phase correlation and relativistic dynamics. For a fully coherent system, the radiation intensity scales quadratically with the number of particles ($N^2$), provided that all charges radiate in phase within the coherence volume defined by the radiation wavelength $\lambda$ and angular frequency $\omega$. However, as the bunch volume exceeds the coherence volume ($\sim \lambda^3$), spatial variations in charge distribution introduce phase mismatches, leading to reduced coherence. As shown in the left panel of Figure 2, we present the different $\sigma_z$ (the bunch size in the $z$-direction) for the spectral-angular distribution of radiation emitted by bunches. An increase in $\sigma_z$ signifies an expansion of the bunch volume beyond the coherence volume, which introduces spatial phase mismatches due to the growing disparity between the bunch dimensions and the characteristic wavelength of the emitted CChR. The right panel of Figure 2 shows that a larger $z_b$ implies a greater tilted angle $\psi$, which deviates from the Cherenkov radiation angle, consequently leading to a reduction in the radiation intensity. By precisely aligning the electron beam angle with the Cherenkov angle, both experimental studies and numerical simulations have consistently demonstrated that the radiation intensity of CChR is maximized (S. Y. Gogolev & A. P. Potylitsyn 2019; Y. Tadenuma et al. 2022). This theoretical framework can also be extended to the magnetized plasma environment within the magnetosphere of a magnetar.

Within the framework of coherent curvature radiation, FRB analogs at frequencies significantly higher than the GHz range are expected to be dimmer due to the reduced coherent volume (W. Lu & P. Kumar 2018). However, in the case of CChR, the luminosity is highly sensitive to the Lorentz factor of the relativistic electrons ($L \propto \gamma_e^8 \nu^{-2}$; Z.-N. Liu et al. 2023a), allowing it to maintain a high level of brightness even at THz frequencies. Thus, the intrinsic emission of FRBs could theoretically extend into the millimeter wavelengths within the CChR framework. In the following section, we will provide a detailed discussion on coherent THz Cherenkov radiation from FRBs and THz-emitting magnetars.

### 3. Coherent THz Cherenkov Radiation from FRBs and THz-emitting Magnetars

Magnetars are highly magnetized neutron stars with extraordinarily strong magnetic fields (R. C. Duncan & C. Thompson 1992). The majority of magnetars are detected primarily at X-ray and gamma-ray wavelengths. Among the 30 magnetars listed in the catalog (see The Magnetar Catalog;[5] S. A. Olausen & V. M. Kaspi 2014), only 6 have had their radio emissions detected (F. Camilo et al. 2006, 2007; L. Levin et al. 2010; R. P. Eatough et al. 2013; C. D. Bochenek et al. 2020; P. Esposito et al. 2020). SGR J1745-2900, one of the radio-emitting magnetars, is particularly fascinating because of its proximity to Sgr A* in the Galactic center (G. C. Bower et al. 2015). Interestingly, P. Torne et al. (2015) detected the magnetar SGR J1745-2900 at frequencies up to 225 GHz, setting a record for pulsar emission at that time. Only very few normal pulsars have been detected above 30 GHz (R. Wielebinski et al. 1993; M. Kramer et al. 1997; D. Morris et al. 1997; O. Löhmer et al. 2008). Subsequently, P. Torne et al. (2017) reported new observations of the same magnetar, extending the detection to frequencies of up to 291 GHz and revealing evidence of high linear polarization in its millimeter-wave emission. The radio emissions of magnetars resemble those of regular pulsars but exhibit several notable distinctions. These include variations in flux density, spectral index, pulse-profile shape, and polarization properties (F. Camilo et al. 2006, 2007; M. Kramer et al. 2007; K. Lazaridis et al. 2008; L. Levin et al. 2012; R. S. Lynch et al. 2015). In the millimeter wavelength range, the most intense pulses exhibited a peak flux density of 19 Jy at 101 GHz, with a duration of 1 ms (P. Torne et al. 2015). This implies an extraordinary brightness temperature exceeding $10^{23}$ K, at a distance of 8.3 kpc from the Galactic center. The peak flux density of the burst corresponds to a luminosity of $1.6 \times 10^{35}$ erg s$^{-1}$ at 101 GHz.

---
[5] https://www.physics.mcgill.ca/~pulsar/magnetar/main.html

Observations of most pulsars indicate that the degree of linear polarization generally decreases with increasing observing frequency (A. G. Lyne et al. 1971; R. N. Manchester 1971; P. A. Hamilton et al. 1977; D. M. Gould & A. G. Lyne 1998; S. Johnston et al. 2006). To explain the polarization characteristics, extensive theoretical research on pulsar polarization has been conducted, focusing on emission processes (e.g., R. X. Xu et al. 2000; J. Dyks et al. 2010) and propagation effects (e.g., C. Wang et al. 2010; V. S. Beskin & A. A. Philippov 2012), both of which can contribute to the depolarization of pulsar linear polarization. The detection of highly polarized emission from SGR J1745-2900 at millimeter wavelengths contrasts sharply with the behavior of typical radio pulsars, which are known to undergo depolarization at higher frequencies (D. Morris et al. 1981; K. M. Xilouris et al. 1996). Other radio magnetars have been observed to retain strong linear polarization even at very high frequencies (M. Kramer et al. 2007; F. Camilo et al. 2008). However, a comprehensive and plausible explanation for this observation is currently lacking.

We briefly discuss the potential of coherent curvature radiation as an explanation for the observed THz-emitting magnetars. Given that magnetic field lines are inherently curved, particles emit radiation as they accelerate along these curved trajectories. The characteristic frequency of curvature radiation can be given by

$$\nu_{\rm CR} = \frac{3}{4\pi}\gamma_e^3 \frac{c}{\rho} = 0.23~{\rm THz}~\gamma_{e,2.5}^3 \rho_6^{-1}, \qquad (2)$$

where $\rho$ is the curvature radius. Although curvature radiation can effectively explain the THz emission from magnetars in terms of frequency, the frequency-dependent behavior of linear polarization presents certain challenges. Specifically, within the framework of coherent curvature radiation, the linear polarization fraction decreases at higher frequencies (A. G. Lyne et al. 1971; R. N. Manchester 1971; P. A. Hamilton et al. 1977; D. M. Gould & A. G. Lyne 1998; S. Johnston et al. 2006; Z.-N. Liu et al. 2023b). Therefore, curvature radiation may struggle to explain the observed high linear polarization in the millimeter-wave emission of SGR J1745-2900 (P. Torne et al. 2017).

Notably, Y. Tadenuma et al. (2022) have confirmed that the coherent THz Cherenkov radiation pulse exhibits nearly linear polarization within the plane defined by the electron beam trajectory and the detector. Besides, the strongly linearly polarized CChR emitted by a single fast particle has been successfully employed to explain the polarization characteristics of the majority of repeating FRBs (Z.-N. Liu et al. 2023a). We consider coherent THz Cherenkov radiation as an explanation for THz-emitting magnetars. High linear polarization is maintained even at high frequencies. As shown in Figure 3, we present the linear polarization as a function of $\psi$ for CChR. Within the radiation cone of CChR, the emitted radiation exhibits nearly 100% linear polarization, regardless of the tilt angle of the bunches. The characteristic frequency in the lab frame can be written as (Z.-N. Liu et al. 2023a)

$$\nu_{\rm CChR} = \frac{\omega_c}{2\pi} = \frac{1}{2\pi}\gamma_p \omega_p \,|\gamma_e \gamma_p (\beta - \beta_p)|^{-1}$$
$$\simeq 0.17~{\rm THz}~\xi_2^{1/2} B_{\rm s,15}^{1/2} P_0^{-1/2} R_7^{-3/2} \gamma_{p,2}^{-1/2} \gamma_{e,2.5}^{-1}, \quad (3)$$

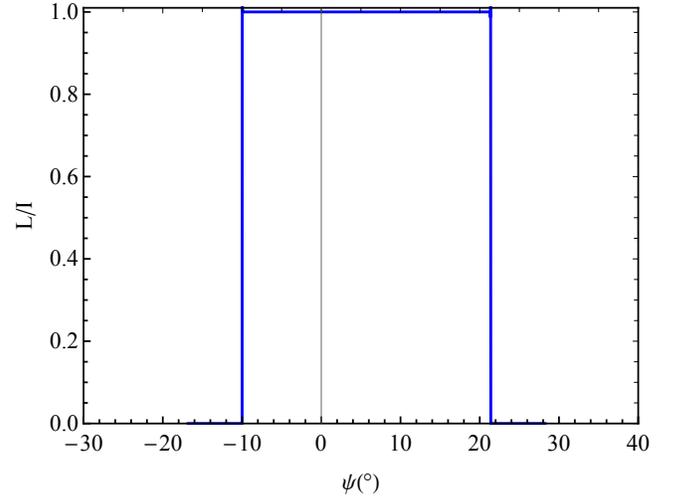

**Figure 3.** The linear polarization as a function of $\psi$ for CChR by tilted bunches. The following physical parameters are employed in our calculation: $\gamma_e = 10^{2.5}$, $\gamma_p = 10^2$, $P = 1$ s, $B_s = 10^{15}$ G, $\xi = 100$.

$\omega_c = \gamma_p \omega'_c$ where $\xi$ represents the multiplicity parameter. The characteristic frequency of CChR can naturally account for the observed THz emission from magnetars (P. Torne et al. 2015, 2017) while also predicting the potential existence of THz counterparts to FRBs. When relativistic charged particles pass through a magnetized plasma, the condition $n_r^2 > 1$ can be met to produce coherent THz Cherenkov radiation. The emitted frequency forms a hollow cone with an opening angle around the trajectory of the emitting particle, where the Cherenkov angle can be given by $\theta_c = \arcsin(1 + (\omega/\omega_c)^2)^{-1/2} \simeq 10^{-1}$ rad. We have calculated the total true luminosity for CChR in the lab frame can be given by (Z.-N. Liu et al. 2023a) $L \simeq N_b(\eta N_e)^2 P_e^{\rm ChR} \gamma_e^2$, where $N_b$ denotes the number of bunches, $N_e$ represents the number of charged particles, and $P_e^{\rm ChR}$ is the emission power of an individual electron for CChR. As shown in Figure 4, we show the relationship between the THz frequency, radiation radius, and luminosity. THz radiation is expected to originate from regions proximal to magnetars, with its luminosity capable of explaining the observed emission from SGR J1745-2900 (P. Torne et al. 2015) while simultaneously providing predictions for potential THz counterparts to FRBs.

The CChR mechanism has successfully explained various observed characteristics of FRBs, including frequency drifting, polarization, and spectrum, etc. (Z.-N. Liu et al. 2023a). In this work, by considering the geometric structure of tilted bunches, we further supplement the explanation for other observational features of FRBs. On average, apparently nonrepeating FRBs have a higher energy $\log(E) = 40.0 \pm 1.1$ and luminosity $\log(L) = 42.6 \pm 1.1$ compared to repeating FRBs, which exhibit a lower energy $\log(E) = 39.5 \pm 1.4$ and luminosity $\log(L) = 41.7 \pm 1.3$ (S.-Q. Zhong et al. 2022). Under the assumption that repeating and nonrepeating FRBs originate from a single population, the coherence of radiation in the CChR mechanism is influenced by the tilt angle of the bunch. This difference in coherence leads to the observed variations in energy and luminosity between repeating and nonrepeating FRBs.

Extensive observations of repeating FRBs, particularly by FAST, have significantly enriched the data for analyzing waiting time distributions. Some studies have consistently revealed a characteristic bimodal distribution, featuring distinct peaks at both millisecond and second timescales (M. Cruces et al. 2021; D. Li et al. 2021; H. Xu et al. 2022; Y.-K. Zhang et al. 2023). The waiting time distribution of FRBs may be associated with their triggering mechanisms, such as magnetar starquakes (Q.-C. Li et al. 2022) or the magnetar–asteroid belts model (Z. G. Dai et al. 2016). However, the millisecond-scale waiting time distribution could additionally be affected by the intrinsic radiation mechanisms of FRBs. Since Cherenkov radiation is not emitted along the relativistic particle's velocity direction ($\theta_c = 0$), when the line of sight sweeps across the Cherenkov radiation cone, two spatially separated emission regions will naturally emerge. This geometry corresponds to clustered FRB substructures and can explain the observed millisecond-scale waiting time distribution in repeating FRBs.

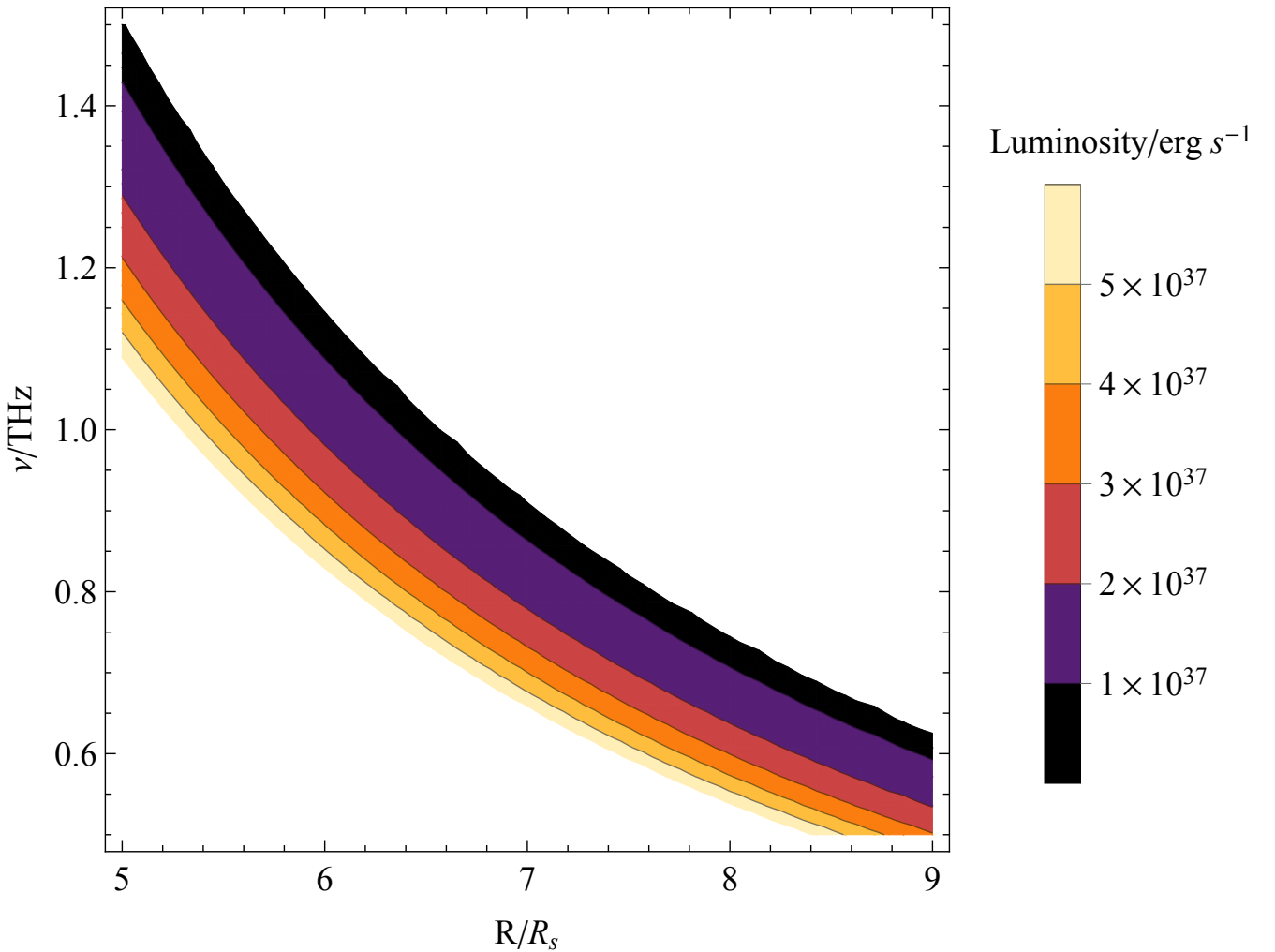

**Figure 4.** Within the framework of CChR in the magnetosphere of a magnetar, the relationship between the THz frequency, radiation radius, and luminosity is presented. The following physical parameters are employed in our calculation: $\gamma_e = 10^{2.5}$, $\gamma_p = 10^2$, $P = 1$ s, $B_s = 10^{15}$ G, $\xi = 100$, stellar radius $R_s = 10^6$ cm.

A periodicity of ∼1.7 s was recently identified in FRB 20201124A (C. Du et al. 2025), showing a clear signal on two particular days. The study suggests that FRB 20201124A could be associated with a young magnetar. However, whether the periodicity observed during these two specific days is coincidental remains controversial. Reflecting on the FRB 200428 event, without prior knowledge that SGR J1935+2154 was a magnetar and observing only FRB 200428 in isolation, we might have missed this landmark discovery entirely. Searching for periodicity at the burst's peak flux density would have been particularly challenging since it was a one-off event. However, periodicity searches at lower flux densities proved effective, as stable radio periodic signals from the magnetar had already been detected at low flux densities before the FRB 200428 event. If this interpretation holds, it implies that the strength of the coherence of FRBs may significantly affect our ability to detect underlying periodicity. In our study, the coherence of FRBs radiation may be correlated with some physical parameters, including the inclination angle of bunches, geometric configuration of tilted bunches, and properties of magnetized plasma. Given this challenge, using flux density as a threshold for burst periodicity searches would prove to be an effective approach. Furthermore, radio pulsations from SGR J1935+2154 were localized to a stable magnetospheric region, while its burst emissions appeared to originate from spatially stochastic locations (W. Zhu et al. 2023). Nonstationary FRB emission regions obscure periodicity, while pulsar pulses from fixed polar caps maintain obvious periodicity. Consequently, the positional uncertainty of FRB emission regions and the coherence of their radiation pose significant challenges to periodicity searches.

Within the framework of CChR, we estimate the fraction of FRBs capable of producing THz counterparts. Equation (3) demonstrates that the radiation frequency of CChR is strongly influenced by three key parameters: the emission radius, the Lorentz factor of relativistic electrons $\gamma_e$, and the Lorentz factor of plasma $\gamma_p$. Based on the calculations presented in Figure 4, we consider a random distribution of the emission radius (ranging from $5 \times 10^6$ to $10^8$ cm) and assume Gaussian distributions for both $\gamma_e$ and $\gamma_p$. Our simulations yield a THz-to-GHz burst ratio of $\sim 0.21$. Notably, the higher propagation efficiency of high-frequency radiation through magnetospheric plasma implies that the observable THz counterpart fraction could be significantly enhanced relative to this intrinsic production ratio.

If both repeating and apparently nonrepeating FRBs originate from the same population, the CChR could provide a unified scenario to explain their observational differences. Within the framework of CChR, the tilt angle of bunches influences the coherence of radiation. When the angle $\psi$ approaches the Cherenkov angle $\theta_c$, the radiation maintains high coherence, making these bursts detectable as repeaters. However, for bursts where $\psi$ significantly deviates from $\theta_c$, the radiation intensity may drop below the detection threshold of telescopes. In such cases, the repeating behavior of these apparently nonrepeaters becomes observationally inaccessible. All FRBs may be intrinsically repeating sources, but their observational classification as repeaters and apparently nonrepeating FRBs hinges on whether the coherence level exceeds the telescope's detection threshold.

## 4. Conclusions and Discussion

In this paper, we have refined the bunched CChR framework by incorporating parameters for bunch inclination and geometric configuration, allowing for a more rigorous modeling of FRB emission and THz radiation from magnetars. When relativistic bunches are injected into the magnetized plasma of a magnetar's magnetosphere at the Cherenkov angle, their emitted waves undergo constructive interference. Additionally, the three-dimensional structure of the bunches plays a critical role in shaping the coherence of the radiation. A larger $z_b$ corresponds to an increased tilt angle, causing deviation from the Cherenkov angle. An increase in $\sigma_z$ indicates an expansion of the bunch volume beyond the coherence volume, leading to spatial phase mismatches. Within the CChR framework, we predicted the presence of THz emission counterparts associated with FRBs and provided an explanation for the observed characteristics of the THz-emitting magnetar SGR J1745-2900. The inclination angle of the bunch has a significant impact on the coherence of radiation. Specifically, when the inclination angle of the bunch deviates from the Cherenkov radiation angle, the radiation intensity decreases markedly. The low-energy FRB-like bursts that occurred after the FRB 200428 event might suggest that the radiation geometry of the bunch deviates from the $\theta_c$. Furthermore, the counterparts of the magnetar SGR J1745-2900 reported in P. Torne et al. (2015) at GHz frequencies could also be FRB-like bursts.

Wave–particle resonance-induced scattering in the nonlinear region of the magnetosphere is excited when (Y.-C. Huang & Z.-G. Dai 2024)

$$\sin\theta < \frac{\omega_B}{\omega} < a_0 \sin\theta, \qquad (4)$$

where $\theta$ is the angle between the propagation direction of a radio pulse and the background magnetic field line, $a_0 = eE_0/(m_e c\omega) \approx 70 L_{\text{iso},37}^{1/2} \nu_9^{-1} r_9^{-1}$ is the strong-wave factor of the pulse, $\omega_B/\omega \approx 3 \times 10^3 B_{s,15} \nu_9^{-1} r_9^{-3}$ is the ratio of the electron cyclotron frequency to the wave frequency, and $r = (10^9 \text{cm}) r_9$ is the radius at which scattering is most severe. For a luminous radio pulse with isotropic luminosity $\sim 10^{37}$ erg s$^{-1}$, the scattering is greatly suppressed, since the initiation condition

$$a_0 \left(\frac{\omega_B}{\omega}\right)^{-1} \sin\theta \sim 10^{-2} L_{\text{iso},37}^{1/2} B_{s,15}^{-1} r_9^2 \sin\theta > 1 \qquad (5)$$

is hard to realize. In the case of THz pulses, the condition presented in Equation (4) becomes more difficult to satisfy. Even if the stringent condition can be satisfied, a small beaming angle can still facilitate the free escape of the pulse from the magnetosphere (Y.-C. Huang & Z.-G. Dai 2024).

At millimeter wavelengths, the brightest millisecond-scale pulses from SGR J1745-2900 reached a peak flux density of 19 Jy at 101 GHz, corresponding to a luminosity of $\simeq 1.6 \times 10^{35}$ erg s$^{-1}$ (P. Torne et al. 2015). Notably, the FRB 200428 burst peak luminosity emitted in the 400–800 MHz is $7_{-4}^{+7} d_{10 \text{ kpc}}^2 \times 10^{36}$ erg s$^{-1}$ (CHIME/FRB Collaboration et al. 2020). The two sources differ in luminosity by roughly 1 order of magnitude, suggesting that the brightest pulses from SGR J1745-2900 could be a missed FRB. Therefore, long-term monitoring of the magnetars (e.g., SGR J1935+2154, SGR J1745-2900), particularly in the millimeter wave band, will be crucial for unraveling the physical origins of both FRBs and magnetars. Very recently, J. Vera-Casanova et al. (2025) observed eight highly polarized pulses originating from the Galactic center magnetar SGR J1745-2900 by utilizing the Atacama Large Millimeter/submillimeter Array (ALMA) for millimeter-wave observations. The cumulative energy distribution of these pulses aligns with the values reported for magnetars at centimeter wavelengths and repeating FRBs, suggesting that THz-emitting magnetars and FRBs may have similar physical origins and radiation mechanisms.

The IRAM 30 m telescope (M. Carter et al. 2012) would theoretically observe the THz emission relevant to FRBs. Although FRBs may emit faint emission in the sub-THz band, the telescope's current detector system is primarily designed for continuum and spectral line observations rather than millisecond-timescale transient detection. However, FRB detections remain feasible during deep-field observations, particularly for sources showing extended or repeating THz emission. The unprecedented sensitivity of upgraded millimeter telescopes like ALMA and IRAM is revealing a substantial population of faint magnetars, providing crucial insights into the radio emission mechanisms of both FRBs and magnetars, shedding light on their potential physical connections.

## Acknowledgments

We are grateful to Bing Zhang, Xue-Feng Wu, Fa-Yin Wang, Yuan-Pei Yang, Yi Feng, Jin-Jun Geng, and an anonymous referee for the helpful discussions and constructive suggestions. This work was supported by the National Natural Science Foundation of China (grant No. 12393812), the National SKA Program of China (grant No. 2020SKA0120302), and the Strategic Priority Research Program of the Chinese Academy of Sciences (grant No. XDB0550300). W.-Y.W. acknowledges support from the NSFC (Nos. 12261141690 and 12403058), the National SKA Program of China (No. 2020SKA0120100), and the Strategic Priority Research Program of the CAS (No. XDB0550300).

## Appendix

As bunches move through a dielectric medium, its electric field polarizes the surrounding plasma, and the resulting polarization current produces electromagnetic radiation. The field $E_b$ of the bunches can be written as (S. Y. Gogolev & A. P. Potylitsyn 2019)

$$E_b(r, \omega) = \{E_{bx}(r, \omega), E_{by}(r, \omega), E_{bz}(r, \omega)\}$$

$$= \frac{Ne\omega}{\pi\gamma(\beta c)^2} \int_{-\infty}^{\infty}\int_{-\infty}^{\infty}\int_{-\infty}^{\infty} g(\psi) e^{i\omega(z-z_b)/\beta c_*}$$

$$\left( \frac{\{x-x_b, y-y_b, 0\} K_1\left(\frac{\omega\sqrt{(x-x_b)^2+(y-y_b)^2}}{\beta c \gamma}\right)}{\sqrt{(x-x_b)^2+(y-y_b)^2}} \right.$$

$$\left. - \frac{i\{0,0,1\} K_0\left(\frac{\omega\sqrt{(x-x_b)^2+(y-y_b)^2}}{\beta c \gamma}\right)}{\gamma} \right) dx_b dy_b dz_b, \quad (A1)$$

where the electric field $E_b$ of a bunch consisting of $N$ electrons with coordinates $r_j = x_j, y_j, z_j$ in the frame connected to the center of the bunch. $K_{0,1}$ are the modified Bessel functions of the second kind. If the "pancake-like" bunch is tilted relative to the OZ axis by an angle $\psi$, the charge distribution of the bunch can be represented as an asymmetric three-dimensional Gaussian (A. P. Potylitsyn 2016; G. Kube & A. P. Potylitsyn 2018)

$$g(\psi) = \frac{e^{-\frac{x_b^2}{2\sigma_x^2} - \frac{(y_b \cos(\psi) + z_b \sin(\psi))^2}{2\sigma_y^2} - \frac{(z_b \cos(\psi) - y_b \sin(\psi))^2}{2\sigma_z^2}}}{(2\pi)^{3/2}\sigma_x\sigma_y\sigma_z}, \quad (A2)$$

where $\sigma_x$, $\sigma_y$, $\sigma_z$ denote the bunch sizes in the respective directions. In the case of a magnetic plasma with dimensions $2H \times 2H \times L$ (See Figure 5), the spectral-angular distribution of radiation emitted by bunches traveling through the magnetic plasma can be given by (S. Y. Gogolev & A. P. Potylitsyn 2019)

$$\frac{d^2W_b}{d\omega d\Omega} = \frac{d^2W_{b\parallel}}{d\omega d\Omega} + \frac{d^2W_{b\perp}}{d\omega d\Omega}, \quad (A3)$$

$$\frac{d^2W_{b\parallel}}{d\omega d\Omega} = \frac{N^2 e^2}{\pi^2 c} \left| \frac{2\pi(e^{\frac{i\omega L(1-\beta\sqrt{\varepsilon-\sin^2(\theta)})}{\beta c}} - 1)(\varepsilon-1)}{\beta\gamma\lambda^2(1-\beta\sqrt{\varepsilon-\sin^2(\theta)})\varepsilon} \right|^2 *$$

$$(\sin^2(\theta) + |\sqrt{\varepsilon-\sin^2(\theta)}|^2) *$$

$$\left| \int_{-H}^{H}\int_{-H}^{H}\int_{-\infty}^{\infty}\int_{-\infty}^{\infty}\int_{-\infty}^{\infty} \frac{\sqrt{\varepsilon}\cos(\theta)g(\psi)e^{-i\Delta\Phi}}{\cos(\theta)+\sqrt{\varepsilon-\sin^2(\theta)}} * \right.$$

$$\left. \frac{(X_b\cos(\phi) - Y_b\sin(\phi)) K_1\left(\frac{\omega\sqrt{X_b^2+Y_b^2}}{c\beta\gamma}\right)}{\sqrt{X_b^2+Y_b^2}} dx dy d^3r_b \right|^2, \quad (A4)$$

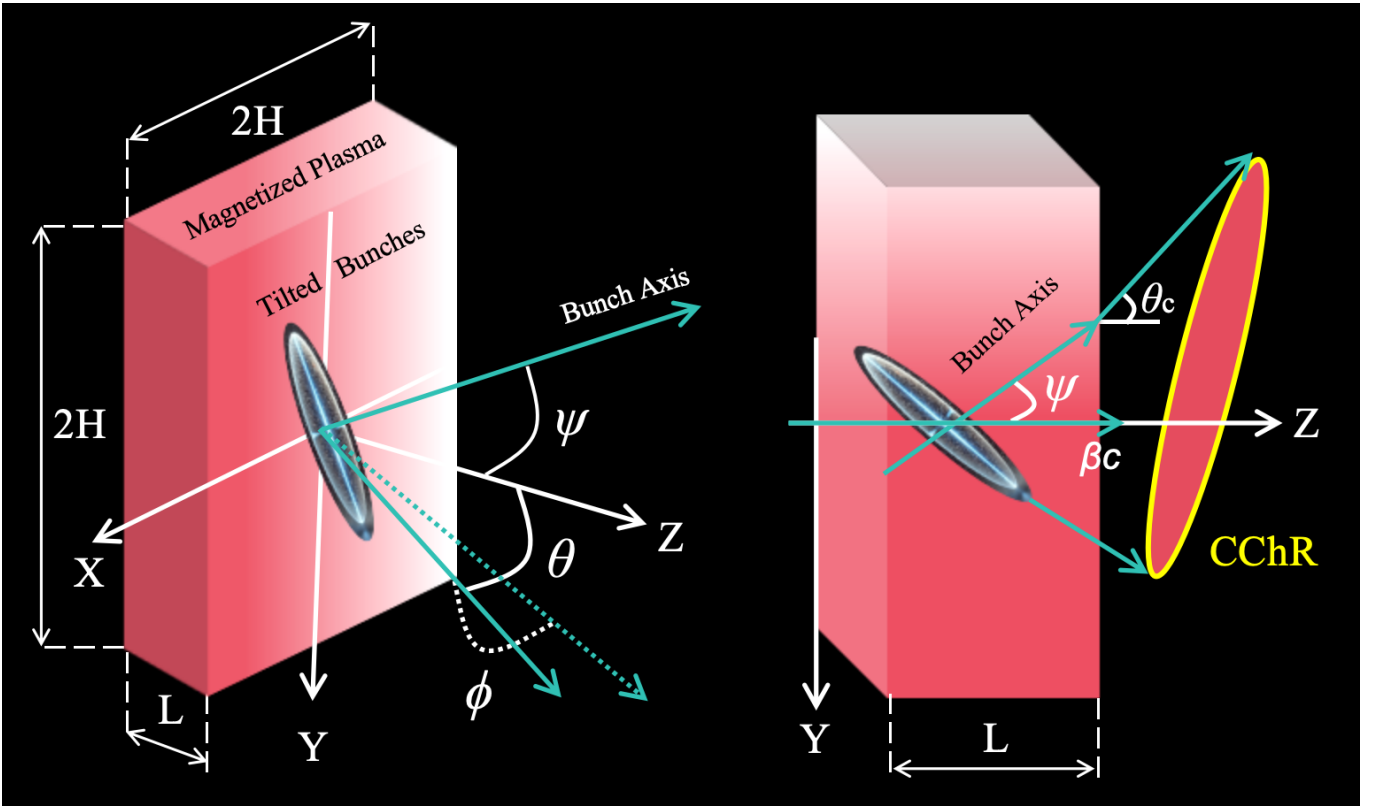

**Figure 5.** Generation of CChR in a magnetized plasma medium with dimensions of $2H \times 2H \times L$, induced by "pancake-like" bunches. $L$ represents the thickness of the medium, $2H$ denotes its transverse size, and $\psi$ signifies the tilt angle of the bunches relative to the medium's axis.

$$\frac{d^2W_{b_\perp}}{d\omega d\Omega} = \frac{N^2 e^2}{\pi^2 c} \left| \frac{2\pi(e^{\frac{i\omega L(1-\beta\sqrt{\varepsilon-\sin^2(\theta)})}{\beta c}} - 1)(\varepsilon - 1)}{\beta\gamma\lambda^2(1 - \beta\sqrt{\varepsilon - \sin^2(\theta)})} \right|^2 *$$

$$\left| \int_{-H}^{H} \int_{-H}^{H} \int_{-\infty}^{\infty} \int_{-\infty}^{\infty} \int_{-\infty}^{\infty} \frac{\cos(\theta) g(\psi) e^{-i\Delta\Phi}}{\varepsilon \cos(\theta) + \sqrt{\varepsilon - \sin^2(\theta)}} * \right.$$

$$\left( \frac{\sqrt{\varepsilon - \sin^2(\theta)}(X_b \sin(\phi) + Y_b \cos(\phi)) K_1\left(\frac{\omega\sqrt{X_b^2 + Y_b^2}}{c\beta\gamma}\right)}{\sqrt{X_b^2 + Y_b^2}} \right.$$

$$\left. \left. + \frac{i \sin(\theta) K_0\left(\frac{\omega\sqrt{X_b^2 + Y_b^2}}{c\beta\gamma}\right)}{\gamma} \right) dx dy d^3 r_b \right|^2, \quad (A5)$$

where $\theta$ and $\phi$ are the angular variables characterizing the radiation in a medium, $\Delta\Phi = \omega/c(x\sin(\theta)\sin(\phi) + y\sin(\theta)\cos(\phi) + z_b/\beta)$, $X_b = x - x_b$, $Y_b = y - y_b$, and $r_b = \{x_b, y_b, z_b\}$.

The Stokes parameters can be given by

$$I = A_X A_X^* + A_Y A_Y^*,$$
$$Q = A_X A_X^* - A_Y A_Y^*,$$
$$U = A_X A_Y^* + A_Y A_X^*,$$
$$V = -i(A_X A_Y^* - A_Y A_X^*), \quad (A6)$$

where $A_X = d^2W_{b_\parallel}/d\omega d\Omega$ and $A_Y = d^2W_{b_\perp}/d\omega d\Omega$. The linear polarization degree can be calculated as $L/I$, where $L = \sqrt{Q^2 + U^2}$ represents the total linearly polarized intensity.

## ORCID iDs

Ze-Nan Liu ● https://orcid.org/0000-0002-5758-1374
Wei-Yang Wang ● https://orcid.org/0000-0001-9036-8543
Yu-Chen Huang ● https://orcid.org/0009-0009-7749-8998
Zi-Gao Dai ● https://orcid.org/0000-0002-7835-8585